# Crystallographic dependence of Field Evaporation Energy Barrier in metals using Field Evaporation Energy Loss Spectroscopy mapping

*François Vurpillot*[*,1], *Constantinos Hatzoglou*[,2], *Benjamin Klaes*[1], *Loic Rousseau*[1], *Jean-Baptiste Maillet*[1], *Ivan Blum*[1], *Baptiste Gault*[3,4], *Alfred Cerezo*[5]

[1]Univ Rouen Normandie, INSA Rouen Normandie, CNRS, Normandie Univ, GPM UMR 6634, F-76000 Rouen, France

[2]Department of Materials Science and Engineering, NTNU, Norwegian University of Science and Technology, Trondheim 7491, Norway

[3]Max-Planck Institut für Eisenforschung GmbH, D-40237, Düsseldorf, Germany.

[4]Department of Materials, Imperial College London, London, SW 2AZ , U.K.

[5] Materials Department, University of Oxford, Parks Rd, U.K.

* Corresponding author: francois.vurpillot@univ-rouen.fr

**Abstract**

Atom probe tomography data is composed of a list of coordinates of the reconstructed atoms in the probed volume. The elemental identity of each atom is derived from time-of-flight mass spectrometry, with no local energetic or chemical information readily available within the mass spectrum. Here, we used a new data processing technique referred to as field evaporation energy loss spectroscopy (FEELS), which analyses the tails of mass peaks. FEELS was used to extract critical energetic parameters that characterize the field evaporation process, which are related to the binding energy of atoms to the surface under intense electrostatic field and dependent of the path followed by the departing atoms during the field evaporation process. We focused our study on different pure face centered cubic metals (Al, Ni, Rh). We demonstrate that the energetic parameters extracted from mass spectra can be mapped in 2D with nanometric resolution. A dependence on the considered crystallographic planes is observed, with sets of planes of low Miller indices showing a lower sensitivity to the intensity of the electric field, which indicates a lower effective attachment energy. The temperature is also an important parameter in particular for Al, which we attribute to an energetic transition between two paths of field evaporation between 25K and 60K close to (002) pole at the specimen's surface. This paper shows that the complex information that can be retrieved from the measured energy loss of surface atoms is important both instrumentally and fundamentally.

# 1. Introduction

Atom probe tomography (APT) provides three-dimensional material microanalysis at the nanometre or sub-nanometre scale (Larson et al. 2013; Lefebvre-Ulrikson, Vurpillot, and Sauvage 2016; Michael K. Miller and Forbes 2014; Gault et al. 2021). At the core of APT is the process of field evaporation, wherein atoms located at the surface are removed from the solid in the form of ions under the influence of an intense surface electric field (Waugh, Boyes, and Southon 1976). Typically, the critical field required for field evaporation, devoid of any thermal activation, falls within the range of 10–60 V/nm. Achieving such intense electric fields is easily accomplished through the classical tip effect. For APT, the material is fashioned into a sharply pointed needle with an end radius (R) ranging from 10 to 100 nm and subjected to a high voltage (V) of a few kV. With careful control of field evaporation through adjustment of V, atoms are liberated individually from the surface, ionized, and accelerated by the electrostatic field away from the surface. The specimen itself acts as an electrostatic lens, projecting ions in a deterministic manner from the surface onto a position-sensitive detector (PSD) located typically 10 cm away from the tip. Elemental identification of the projected ion is achieved through time-of-flight mass spectrometry (TOF-MS). Atoms are field evaporated by the superimposition of ultrashort voltage pulses (ns) or laser pulses (ps or fs pulse duration) onto the DC voltage, providing a start time for ions. The impact of the ions on the detector determines the stop time. Post-processing algorithms are employed to associate to each mass-to-charge an elemental nature, and by recalculating the original positions of the detected ions assuming a known projection law (De Geuser and Gault 2017), reconstruct a 3D point cloud with near-atomic resolution (De Geuser and Gault 2020).

However, the resulting 3D volume provides limited information about the chemical nature of the bonds, i.e. metallic, covalent, or ionic, between atoms prior to their departure from the surface. The nature of chemical bonding remains usually imperceptible from studies of the mass spectra, which is a histogram of the counts at a given mass over charge ratio. While extensive research has explored the potential of field evaporation to directly and locally measure the binding energy of single atoms, practical challenges persist in accurately measuring local surface fields due to surface self-reorganization under field evaporation. Additionally, intricate electrostatic devices are required to measure precise binding energies linked to local atomic environments, local fields, and crystallographic locations on the specimen surface(Ernst 1979). More recent work has also postulated that traces of the bonding mechanism can be retrieved from the analysis of the rate of multiple ion field evaporation in APT (Zhu et al. 2018). These studies have proposed assessing the occurrence of multiple field evaporation events to infer bonding mechanisms in thermoelectric and phase-change materials. However, retrieving local physical and chemical properties at the nanoscale in 3D remains challenging, and the details of fine scale energetics within each family is for the moment out of reach. It is worth noting that access to this information is also of fundamental importance for the practical operation of APT. Variations in both the nature of a surface atom and its local neighborhood significantly impact the field evaporation behavior of the material (Ge et al. 1999; Tsong 1978). Analyzing alloys with atoms of diverse chemical nature can prove arduous, necessitating precise tuning of the analysis conditions to avoid species-specific losses for instance (M K Miller 1981; Yamaguchi, Takahashi, and Kawakami 2009; Takahashi, Kawakami, and Kobayashi 2011; Jin et al. 2022; Chang et al. 2019). This underlines the intricate interplay between the physical and chemical nature of the local zone under investigation and the field evaporation process.

In a recent study (Rousseau et al. 2023), we demonstrated that additional information relative to the local binding energy of the field evaporating atoms could be extracted from the shapes of the mass peaks in atom probes operated in voltage pulsing mode and not equipped with energy compensation devices. This method, termed "Field Evaporation Energy Loss Spectrometry" (FEELS), not only aids in identifying and quantifying elements in the analysis but also encodes information regarding atom binding with its neighborhood and energetic aspects related to the field evaporation process (Rousseau et al. 2023). We showcased on various examples how information related to the chemical and structural nature of evaporated atoms could be relatively easily extracted from APT datasets with nanometer spatial resolution.

FEELS relies on the measurement, in logarithmic scale, of the slope of the mass peak tails to extract an energetic parameter, C, consistent with a relative electric field sensitivity. It was found, in good agreement with density functional theory, that this term could be significantly different at grain boundaries for instance in pure Al. The measurement of this term for the same element but in different phases highlighted the possibility to map the local environment of atoms in term of chemical composition or structural properties. However, the deeper physical meaning of C is still unclear, indeed this energetic parameter is linked not only to the binding energies of atoms, but also to the local path that is followed by these atoms in the first stages of the field evaporation process from the surface (Waugh, Boyes, and Southon 1976; Ashton et al. 2020; Ohnuma 2021; Schmidt, Ernst, and Suchorski 1993). From DFT calculation it was found that the energy barrier height variation was different when considering an evaporation path including the roll-up motion on surrounding neighbors before being expelled from the surface compared to a direct lift-out from the surface. As a result, predicting the field evaporation behavior is extremely complicated since the local path is strongly dependent on the local atomic configuration, which is also related to the crystallography, and possibly the temperature at which the field evaporation proceeds.

Here, we focused our attention on pure FCC metals to evaluate experimentally the dependency of C with physical and instrumental parameters. We have measured C as a function of the crystallographic location of atoms on nickel, rhodium and aluminum (Ni, Rh, and Al) varying base temperature and detection rate. The influence of the rolling-up on C has also been explored.

**2. Materials and methods**

*2.1 Experiments on the LEAP 5000XS*

Ni and Al APT specimens were prepared using electrochemical polishing, starting from pure metal wire specimens and using the solutions and conditions reported in Refs. (Larson et al. 2013; Lefebvre-Ulrikson, Vurpillot, et Sauvage 2016; Michael K. Miller et Forbes 2014; Gault et al. 2021) . APT specimens from Rh were prepared by using a HELIOS 5 dual-beam Xe plasma scanning electron microscope/focused ion beam (SEM/FIB) via classical lift-out specimen preparation process. At the start of the analysis, all specimens possessed a curvature radius of less than 50 nm. Needle-shaped specimens were analyzed using the voltage pulsing mode on a local electrode atom probe with a direct flight path (LEAP 5000XS). Analysis temperatures varied between 25K and 60K for Al and Ni specimens. Rh specimens exhibited greater brittleness, thus analyses were successfully conducted only at 80K in this study. The pulse fraction (PF) was consistently set to 20% across all analyses. During APT specimen evaporation, the detection rate (DR) was maintained constant, with no restrictions on the gradually increasing DC voltage ($V_{DC}$) to offset the progressive blunting of the specimen radius.

*2.2 C value measurement*

In APT, the probability for thermally-activated field evaporation follows an Arrhenius law $K_{EV}$ (Eq. (1)) that is dependent on the energy barrier Q(F) and the temperature T of analysis.

$$K_{EV} \propto exp\left(-\frac{Q(F)}{k_B T}\right) \quad (1)$$

where $k_B$ the Boltzmann constant and $T$ is the temperature, and the activation energy barrier $Q(F)$ depends on an external electric field (F), which can be approximated using a linear approximation:

$$Q(F) = C\left(1 - \frac{F}{F_{EV}}\right) \quad (2)$$

where $F_{EV}$ is the evaporation field of the considered atom, i.e. the field at which the energy barrier is reduced to zero, and $C$ the abovementioned energetic parameter. Non-linearities of Q(F) linked to the path of the adatom at the surface during the escaping process exist (Waugh, Boyes, and Southon 1976; Ashton et al. 2020), but can typically be considered negligible at low temperature and high field (Wada 1984), i.e. the conditions used in the experiments we performed voltage pulsing mode.

The atom is assumed to leave the surface at the top of the pulse and the formed ion acquires almost instantaneously all the kinetic energy induced by the potential energy. The experimental mass-to-charge ratio (M = m/n) is determined using the conservation of energy:

$$M = 2eV_T \left(\frac{t_f^2}{L^2}\right) \quad (3)$$

where $t_f$ is the ion's time of flight, L is the flight distance, e the elementary charge, and $V_T$ the maximum voltage from the DC and pulsed high voltages. Due to the probabilistic nature of the field evaporation process (eq.1 ), some ions can leave the specimen slightly before or after the top of the pulse, some ions hence have an actual energy deficit δ (V=$V_T$(1-δ)), leading to an increase of their time-of-flight, that creates a tail in the mass spectrum. For a mass-to-charge ratio of the evaporated element $M_0$, and to a first order approximation, energy deficit δ can be derived from:

$$M \sim M_0 (1 + \delta) \quad (4)$$

allowing the field at which the ion was field evaporated to be extracted from the measured mass to charge ratio. Indeed, for a certain deficit, the number of evaporated atoms is proportional to Eq. 1, with Q(F) derived from Eq 2. In the mass spectrum, the number of ions at a mass-to-charge M, I(M), can be seen as linearly dependent on δ, and C can be written as:

$$C = k_B T \frac{d(\ln K_{EV})}{d(F/F_{EV})} \sim -k_B T \frac{d(\ln I(M))}{d(M/M_0)} \quad (5)$$

In practice, a peak is chosen from an APT mass spectrum plotted in log scale, and the slope of the mass peak tail is -C/$k_B$T. An illustration of this process is provided in Figure 1a, using an aluminum specimen evaporated in voltage pulse mode on a LEAP 5000X at T=25K. Approximately $5.10^6$ atoms were collected with a DC voltage variation of less than 10%. To accurately measure the C value, a statistically significant number of atoms must be assessed within the mass bins of the mass spectrum. Note that a more correct expression of the shape of the mass peak as a function of δ was found both analytically and numerically. It was found that the linear approximation was sufficient for M>10, above which dynamic effects could be neglected, and for range of mass M>1.005 $M_0$. More details can be found in (Rousseau et al. 2023) .

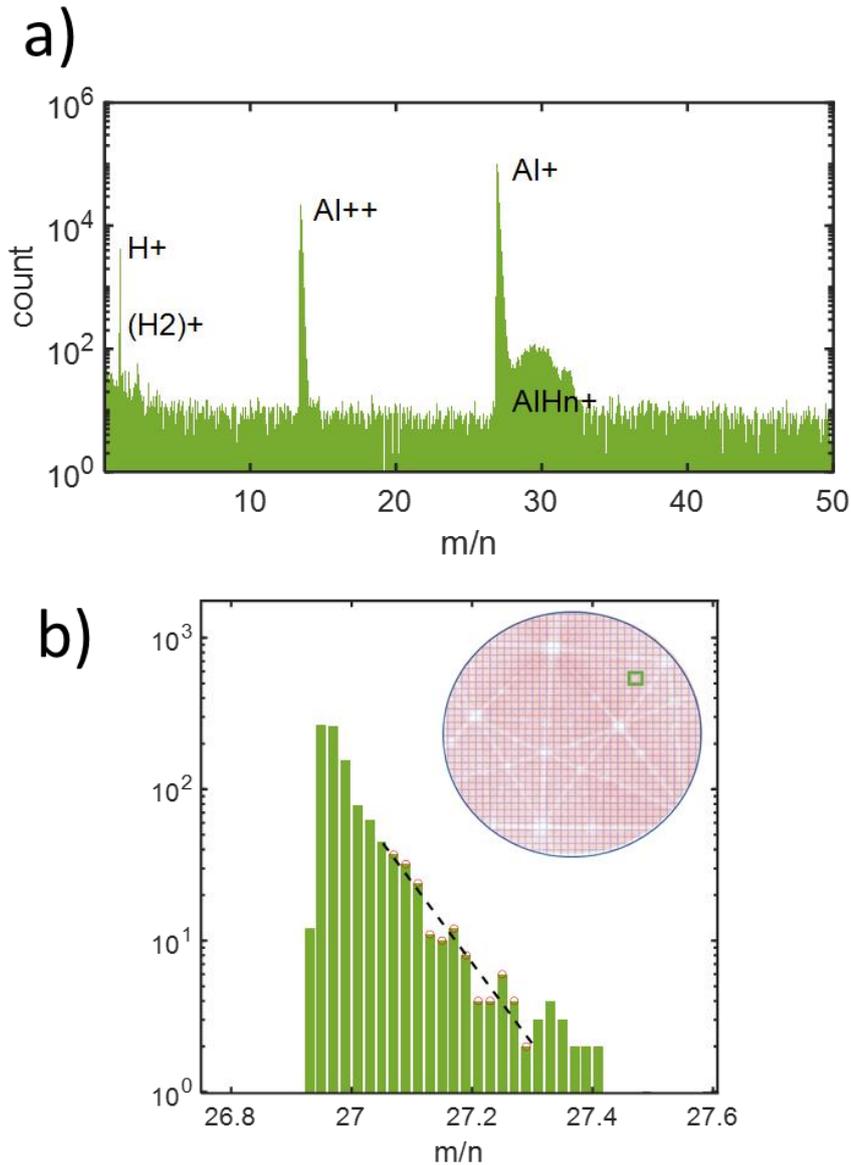

**Fig. 1**: Method used for the determination of the C-value over the detector. a) Example of the mass spectrum of an Al analysis in voltage pulse mode showing main Al+ peak of interest b) zoom in log scale of the Al+ m/n peak (in Da/z) which is extracted locally on the detector map (green square). The peak contains more than $10^3$ atom count. The region situated between 27.1 and 27.3 was selected to calculate the slope of the tail in log scale.

Provided that a sufficient number of ions collected, mass spectra can be calculated from selected locations across the field-of-view and local measurements of C can be performed. Typically, a minimal 1000 to 10000 ions in the considered selected peak are deemed adequate to perform the linear regression, falling between 1.005 and 1.015 $M_0$ in the mass spectrum. This process has been automated under MATLAB™ by discretizing the hit map into a regular mesh, with a spatial binning selected to accommodate the necessary local number of ions. As depicted in Figure 1b, the local C value can be calculated as $C(x,y) = -k_B T \times a(x,y)$, where $a(x,y)$ denotes the slope obtained from the linear regression, and T represents the specimen temperature. Note that the detector hit map shown inset displays the characteristic pattern of poles and zone lines that form in the analysis of many crystalline materials, and is due to crystallographic induced trajectory aberrations and faceting of the specimen around sets of crystallographic planes (Gault et al. 2012; Vurpillot and Oberdorfer 2015). This approach facilitates the generation of a C-map, as detailed in the subsequent section.

## 3. Results

*3.1 Aluminum experiments*

A pure-Al specimen was analyzed at 60K at a DR of 0.5 ions/100 pulses, and a pulse fraction of 20%. A subset of 10 million ions was extracted across a moderate DC voltage increase (less than 5%). Figure 2a plots the map of the density of impacts on the detector or hitmap, which highlights the crystallographic nature of the FCC aluminum specimen. This image was compared to a modelled atomic surface of a sphere intercepting a FCC structure in figure 2b, with indexed main poles. Using the mass spectrum (similar to figure 1), the local fraction of $Al^{++}$ (figure 2c) to the total number of hits, and measured local fraction of Hydrogen (figure 2d) are displayed. These maps are compared to the C-values measured locally on the detector (figure 2e, C-map). C is ranging from 0.6 to 1.2 eV, with apparent local contrasts. The histogram of the C-values across the entire detector, using an adjusted Jet colormap, with C ranging from 0.6 to 1.2 eV is given in figure 2f. Notably, this histogram displays a bimodal behavior, featuring main peaks at approximately 0.65 eV and 0.95 eV, with an average value near 0.85 eV. The corresponding C-map, with the same colormap, is shown inset. It is noteworthy that regions with the lowest hydrogen fraction coincide with regions of lowest $Al^{++}$ fraction, specifically the (002) and (111) regions, indicating a lower electric field (Kingham 1982; Ernst 1979).

In Figure 2c, we report the local radii measured using the Drechsler method on the (002), (113), and (022) poles based on the rings imaged (Drechsler and Wolf 1958; Gault et al. 2008), superimposed over an overlay of the C-map and the hit impact density.

In Figure 2b and 2c are displayed respectively the maps of the locally measured composition of H (maximum of 2% in red) and the local composition in $Al^{++}$ (maximum of 8% in red).

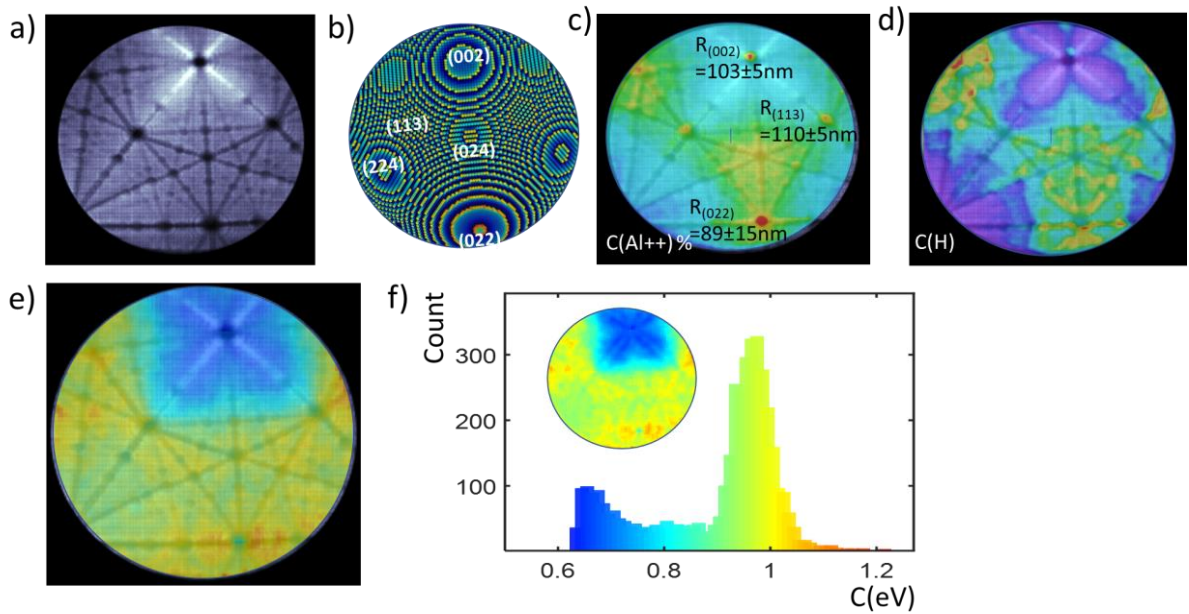

**Fig. 2**: a) map of the density of hit on the detector (hitmap) measured from an Al mass spectrum performed at 60 K, in grayscale (white is high density) of impact b) representation of the atomic surface of the same zone of analysis with main crystallographic poles c) overlay of Al++ map, measured from the same mass spectrum (Jet colorscale with red=8%), with the hitmap d) overlay of H map (H+ , and $H_2+$ peaks), measured from the same mass spectrum (colorscale with red 2%), with the hitmap e) Overlay of the C-value map with the hitmap. The color scale (Jet color scale) is the C-value scale (from 0.5 to 1.3 eV) f) histogram of the C-value of figure e) (Al mass spectrum performed at 60 K), with inset showing the spatial distribution. The color scale is the C-value scale (from 0.5 to 1.3 eV)

A similar analysis was performed for APT data obtained from the very same specimen but at a base temperature of 25K under identical analysis conditions, as reported in Figure 3 a–e. In Figure 3a, the hitmap is reported, Figure 3b reports the Al++ composition map, the lowest Al++ composition is around the (111) pole, just outside the field-of-view. This demonstrates the change in the faceting of the specimen observed previously at different temperatures (Nakamura and Kuroda 1969; Gault et al. 2010) likely which had been attributed to changes in the field evaporation behavior of different sets of atomic planes (Chen and Seidman 1971). Comparatively, the radius of curvature at the (002) pole is slightly lower than that observed in the 60K experiment, decreasing from 103 to 91 nm, whereas the curvature of the (113) poles remains similar. The H composition, Figure 3c, increases slightly at lower temperature, with the lowest values near low-index poles (002) and (111). The C-map (Figure 3d) is now less contrasted with values varying from 0.4 to 0.8 eV, but with fine scale contrasts corresponding to the crystallography of the surface. Note the lowest values at the exact locations of poles centers. Histogram of the C-value (Figure 3e) now exhibits a globally monomodal distribution with an average value of 0.6 eV. The spatial distribution illustrates crystallographic variations ranging from 0.75 eV around the (022) pole to 0.5 eV near the center of the (002) pole. The H fraction, Figure 2b, increases slightly at lower temperature, with the lowest values near low-index poles (002) and (111) in Figure 2c.

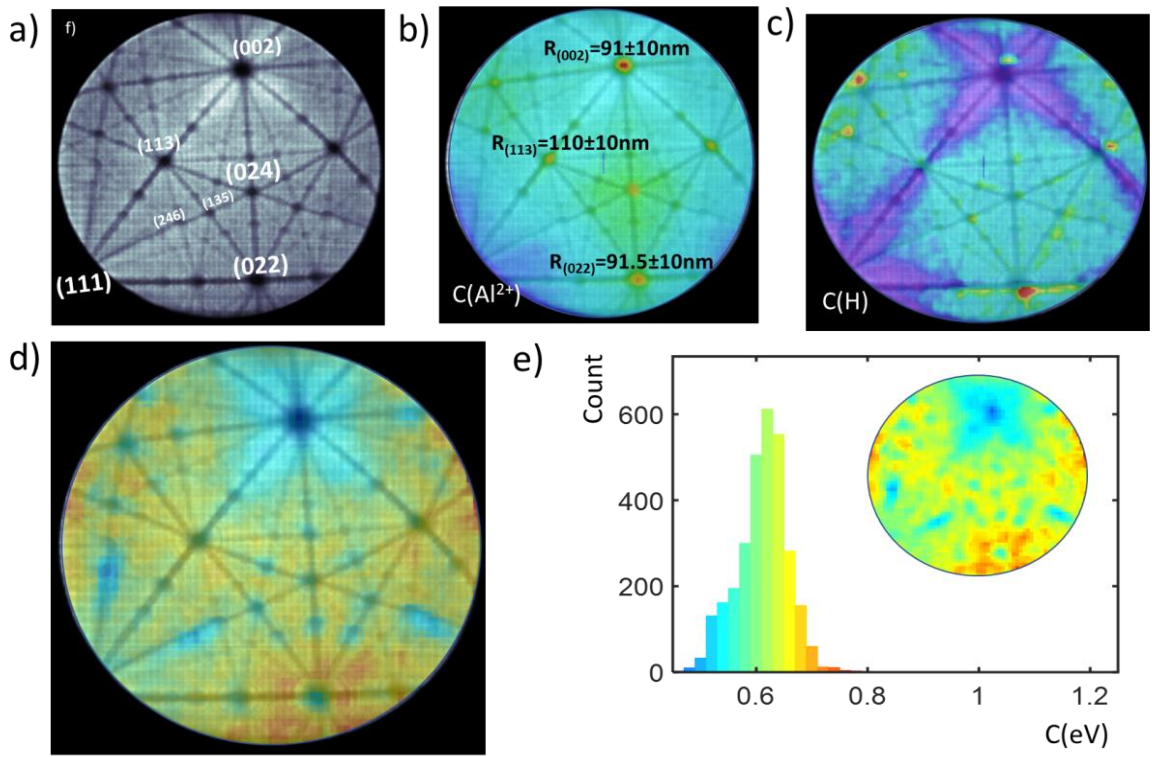

**Fig. 3**: a a) map of the density of hit on the detector (hitmap) measured from an Al mass spectrum performed at 25 K, in grayscale (white is high density) b) overlay of Al++ map, measured from the same mass spectrum (Jet colorscale with red=20%), with the hitmap c) overlay of H map (H+ and $H_2^+$ peaks), measured from the same mass spectrum (colorscale with red 8%), with the hitmap e) Overlay of the C-value map with the hitmap. The color scale (Jet color scale) is the C-value scale (from 0.4 to 0.8 eV) f) histogram of the C-value of figure e) with inset showing the spatial distribution.

*3.2 Rhodium*

In Figure 4, the same method is applied to the analysis of Rh. A selection of 3 million extracted from the analysis of a pure-Rh specimen conducted at 80K, a DR of 5 ions/1000 pulses and a PF of 20% is presented. Figure 4a is the Rh mass spectrum centered on the main peak used for measurement (Rh++). Note that rhodium is principally detected in the 2+ charge state (local field measurement using post-ionization is not possible). Figure 4b illustrates the hit map density, while Figure 4c presents the overlay of the C-map and hit map. A notable correlation between the C-value and the crystallographic structure of the specimen surface is observed, with higher values consistently associated with zone lines and areas having a low density of impacts. Exact centres of (111) and (002) poles are found to have low C-values. Figure 4d depicts the histogram of C-values obtained from Figure 4c. The average C-value is 1.55 eV, with a range of 1.2 to 1.9 eV. Note that the correlation of the C-map with local radii of curvature is here difficult since the measurement is quite imprecise especially on the (022) pole which appears asymmetric. The average values are here similar between poles (i.e. 65±5 nm on the (111) pole, 72±10 nm on the (022) pole, and 70±5 nm on the (002) pole).

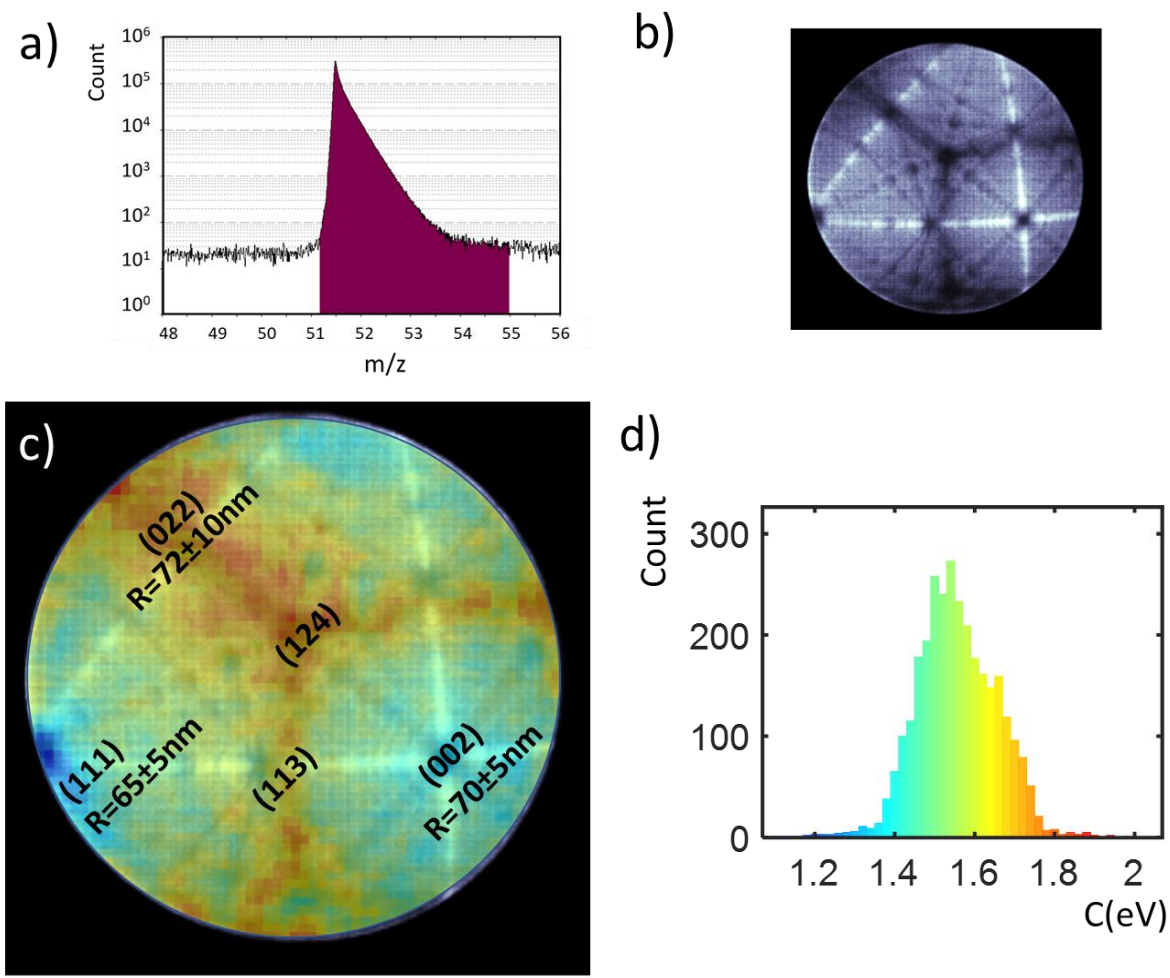

**Fig. 4**: a) mass spectrum of a Rhodium specimen at 80K b) map of the density of hits on the detector (hitmap) c) C-value distribution with a Jet colormap on the detector overlaid with the hitmap. C-value varies between 1.1 eV (blue) an 1.9 eV (red), poles index and measured radius of curvature are indicated in the image d) histogram of the C-value measured on a Rh mass spectrum performed at 80 K. The color scale is the C-value scale (from 1.1 to 1.9 eV).

*3.3 Nickel*

A pure-Ni specimen was analyzed at 25K and 60K, and Figure 5 summarizes the mapping of the C-values in both cases. Specifically, Figure 5a (and Figure 5c) depicts the C-map at 60K (and 25K), yielding average values of 1 eV (and 0.65 eV) respectively. Figures are overlaid with hitmap to highlight crystallographic structures. Additionally, Figure 2b (and Figure 2d) represents the hit map histogram for the 60K analysis (respectively 25K), showing the relatively low dispersion of C-values, with distributions at 60K and 25K having the same global shape. C-maps obtained for two DR, namely 1 ion / 100 pulses and 5 ions / 100 pulses, are plotted in Figure 5a and 5f, respectively, show no significant differences. At both temperatures, and both DR, the highest C-values are consistently observed on high-index poles.

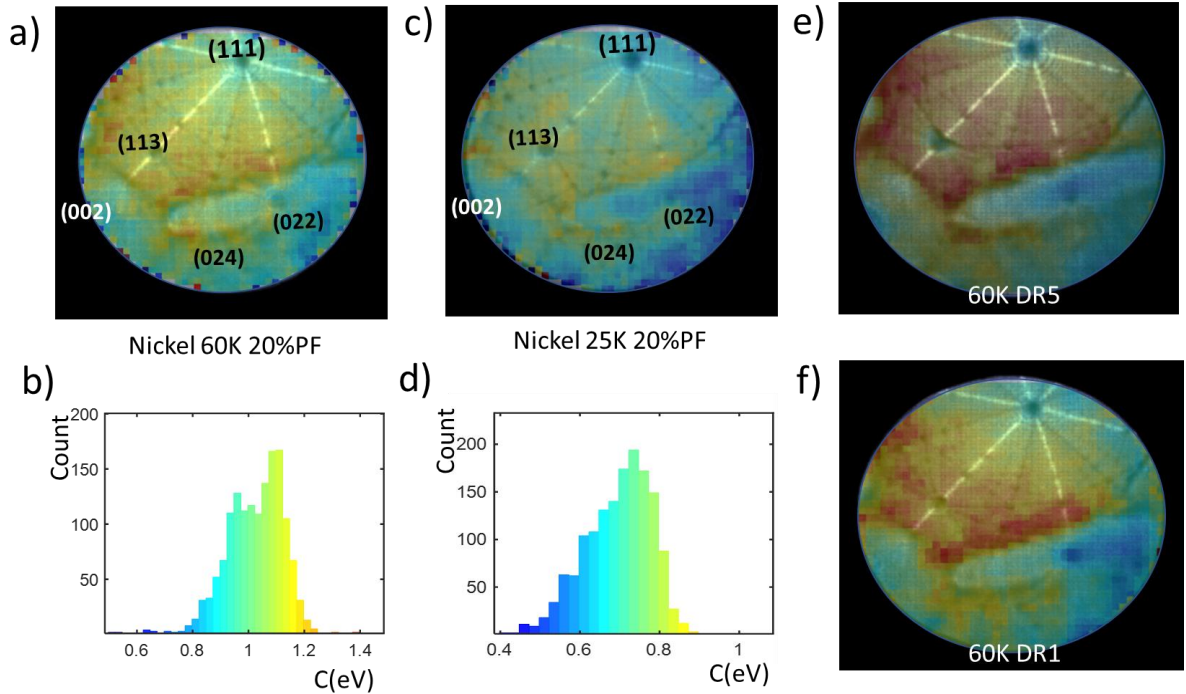

**Fig. 5**: a) overlay of the C-value map of Ni at 60 K, with the hitmap density) The color scale is the C-value scale (from 0.5 to 1.5 eV) b) histogram of the C-value measured on an Ni mass spectrum performed at 60 K c) overlay of the C-value map of Ni at 25 K, with the corresponding hitmap density  The color scale is the C-value scale (from 0.4 to 1.0 eV) c) histogram of the C-value measured on an Ni mass spectrum performed at 60 K e) Overlay of the C-value map of Ni at 60 K, with the hitmap density plotted in 5b, at a detection rate (DR) of 1 /100 pulses and f) 5/100 pulses. (Color scale from a narrower energy range, 0.7 to 1.2 eV to highlight possible contrast difference)

## 4. Discussion

The three FCC materials studied in this paper have common characteristic features in term of the C value. Lowest value of C is always found at the centre of crystallographic poles, and generally the highest values of C are found in the vicinity of high index poles. The average values of C are consistent with previous experimental and computational reports. In Al, an average C value of 0.5 ± 0.1 eV was measured by using the Ernst–Kellogg procedure, i.e. through measurements of the evaporation rate at different base temperature and applied fields (Rousseau et al. 2023). Experiments of Ernst on Rh enables to derive a C-value of 1.5 (±0.5) eV (Ernst 1979), and recent experiment on Ni alloy by Hatzoglou et al. reported a value below 1.1 eV, depending of the composition of the alloy [10].  Here, we observe a large crystallographic dependence, with variations of about 30% between poles in the best case.

We have observed that there is no obvious correlation between the C-value and the surface field, as assessed based on the fraction of $Al^{++}$, which is strongly correlated with the magnitude of the surface electric field according to the post-ionization theory (Kingham 1982). Overall, the C-value tends to increase with the electric field strength (F). Interestingly, regions with the highest electric field, e.g.

near the (024) poles in Al, exhibit relatively lower C-values compared to regions with a slightly lower electric field strength, e.g. near the (022) poles. We have also attempted to correlate the presence of hydrogen with the C-map, but the relationship appears to be complex and not easily interpretable (figure 2d and figure 3c).

For all three metals, we have observed similar trends in the variation of the C-value with temperature. Lower specimen temperatures correspond to lower C-values. For example, the $C_{Al}$ value varies from 1 to 0.5 eV as the temperature decreases from 60 to 25K, while the $C_{Ni}$ changes from 1 eV to 0.65 eV. Although complete data for Rh at 30K were not available, partial measurements (not presented here) suggest a similar decrease. This behavior could be attributed to the relative non-linearity of Q as a function of F at higher temperature, previously reported for a range of metals (Wada 1984). In our approach, we assume a first-order linear expression of Q(F), which is a simplification as confirmed by experimental observations (Wada 1984). This behavior is also evident in density functional theory (DFT) simulations (Ashton et al. 2020)(Ohnuma 2021). In previous works by Ernst and Kellogg (Ernst 1979; Kellogg 1984), the Q(F) function consistently exhibited an increased slope at low field values.

The variation of C with temperature can be explained based on the simplified model illustrated in Figure 7a. The variation of Q with F is represented by the solid black curve. At low temperatures (25K), the maximum pulse reaches a field F(25K), resulting in a lowered energy barrier to $Q(F)_{25K}$ – for Al this is estimated to be 25 meV. Note that this value obtained from of evaporation rate variation with temperature was taken from (Rousseau et al. 2023). The C-value is extrapolated from the slope of Q(F) with respect to F, where energy deficits provide access to ions emitted at lower electric field strengths. The mass peak tail arises from a field variation of 0.5 to 2% below the maximum field. Conversely, at 60K, to maintain the same evaporation rate, the energy barrier must be higher, since Q/kT is constant. Consequently, the electric field is slightly reduced, resulting in an estimated energy barrier of 60 meV in Al at F(60K). At this position, the derivative of Q(F) is more pronounced, leading to a higher C-value.

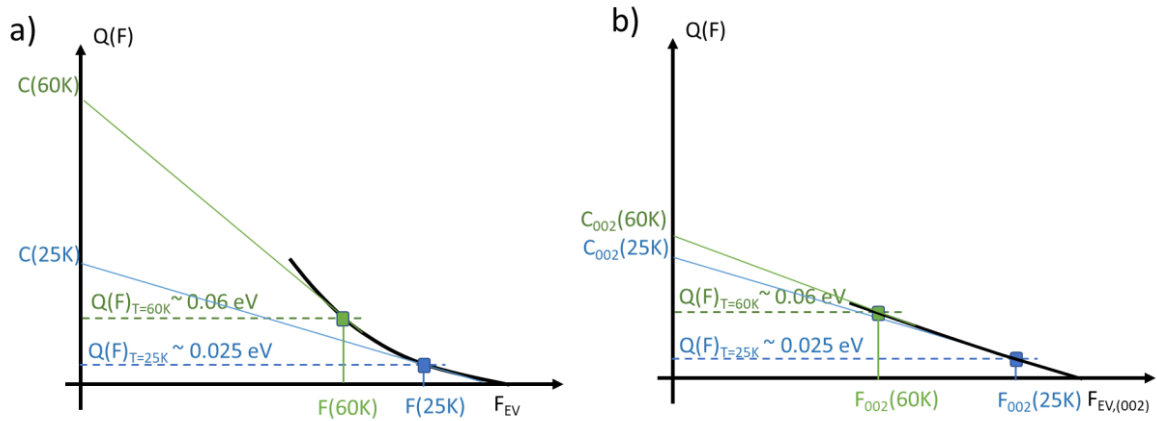

**Fig. 7**: (a) Graphical representation of the variation of the evaporation barrier Q as a function of F, interpreting the measured evolution of the C value at 25K and 60K in the Al case. At 25K, Q~0.025 eV whereas at 60 K, Q=0.060 eV, with a different local slope of Q(F) (b) Same representation but for the local (002) pole region. A more linear variation is proposed to give rise to the C-map contrast observed in figure 2 and 3.

It is interesting to note that this behavior is not systematic over the whole surface in Al. Looking at the histogram of the C-values in Figure 2f and 3e, the lowest peak at about 0.4–0.7 eV is present in both distributions. However, the average value of this peak does not vary strongly with T (from 0.65 to 0.5 eV). This peak comes from the region around the (002) pole. In this case, the variation of Q(F) can be understood through the case illustrated in Figure 7b. In this case, the slope is nearly constant with F, resulting in reduced variations of C with T, $C_{002}(25K) \approx C_{002}(60K)$.

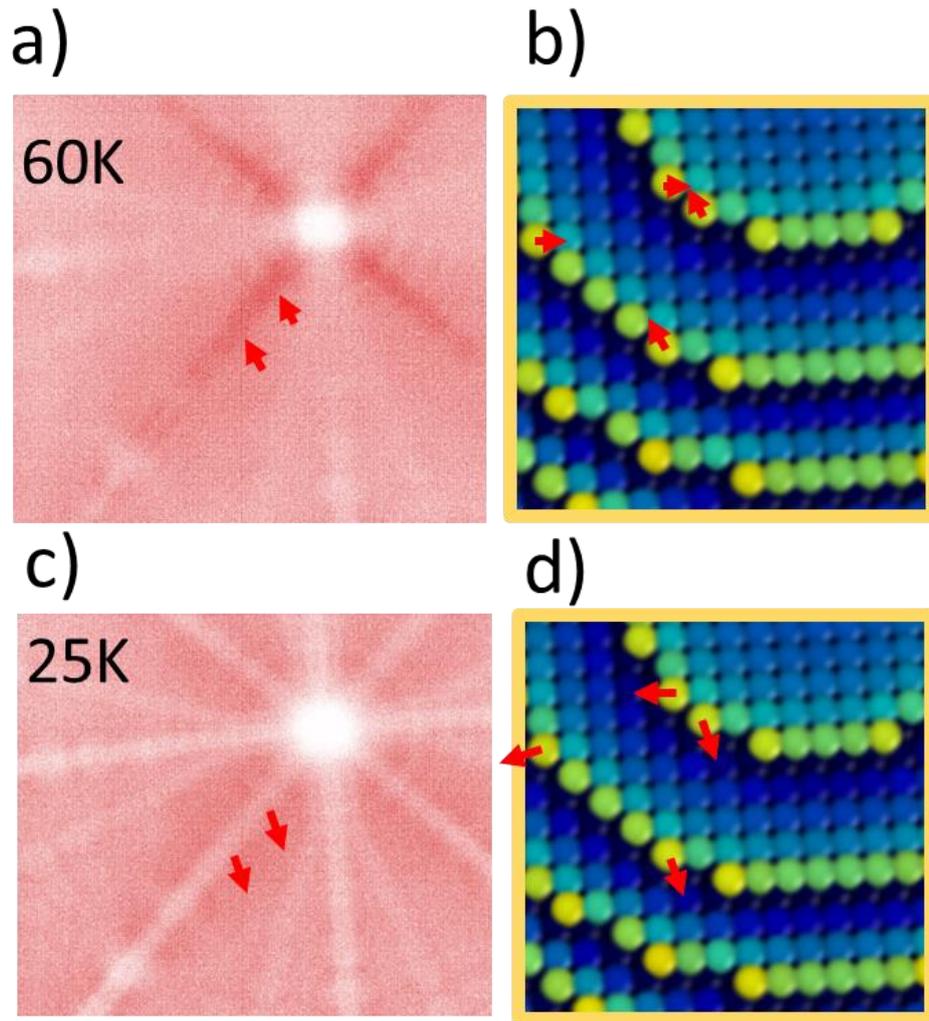

**Fig. 8**: Local hitmap around (002) pole in Al at 60 K(a) and 25 K (c) which is related to the directional walk or deviation of ions in the first step of flight. The rolling effect appear at higher temperature (above 40 K).

The observed behavior is particularly intriguing in the case of the Al specimen, specifically within the vicinity of the (002) pole region. We could argue that this phenomenon is related to the relatively high amount of hydrogen detected in APT analysis for Al, which is different at 25K and 60K. H is known to lower the field evaporation of metals and H is indeed only detected as trace for Ni and Rh (<1%). However, the H composition map presented in figures 2d) and 3c) are relatively similar, whereas C maps evolve strongly between the two cases (especially looking around the (002) pole region). Alternatively, we have correlated this phenomenon with the hitmaps presented in Figure 2a and Figure 3a, focusing on the (002) pole region. While the characteristics of other regions remained consistent, the (002) region exhibited notable changes with decreasing temperature, as detailed Figure 8. Figure 8a is a section of the detector hitmap obtained at 60K, along with, in Figure 8b, a section of the ball model showing the edge of a set of (002) planes. Figure 8c–d are similar but at 25K. At 60K, the zone line between the (002) and (113) poles appears dense, contrasting with its depleted appearance at 25K. These observations are attributed to trajectory artifacts occurring during the initial flight steps (Waugh, Boyes, and Southon 1976; Vurpillot and Oberdorfer 2015). A dense line indicates ions moving inward toward the zone line, as indicated by the red arrows in Figure 8a–b, while a depleted line implies ions moving away from the zone line, Figure 8c–d. The Al desorption hitmap is effectively interpreted through electrodynamic simulations of the field evaporation process. The presence of depleted zone lines can be attributed to the slight repulsion of ions during the initial flight steps, induced by the local electric charges of neighboring atoms at the terrace edges (as observed in Figures 8b and 8d). Above 60K, a transition in the contrast of some zone lines is observed, indicative of a rolling-up motion of the departing atoms. This suggests that Al atoms move over their neighbors before lifting off from the surface, as predicted by DFT (Sanchez, Lozovoi, and Alavi 2004; Ashton et al. 2020). Consequently, the energetic landscape experienced by these atoms differs significantly. While other surface atoms experience an increase in the energy barrier for field evaporation, atoms from the (002) pole follow a smoother path, resulting in a reduced increase in the energy barrier for field evaporation and hence, a lower C-value.

**Conclusion**

In this paper, we present the application of FEELS to three FCC pure metals, namely Al, Ni, and Rh. The asymptotic energy barrier at zero electric field, referred to as the C-value, is deduced from the tails of mass peaks in the mass spectrum obtained in voltage pulsed atom probe without energy compensation devices. We were able to reveal an influence of the crystallography of the specimen surface on this C-value, likely associated to the local atomic arrangements at the surface, with low index poles, i.e. more closed-packed planes, exhibiting a lower C-value. However, at low temperatures, the slope of the energy barrier with field remains relatively similar (within 30-40%) as a function of crystallography.

The local variation of the C-value as a function of specimen temperature was measured to be stronger in Al than for Ni and Rh. This variation could be attributed to the rolling-up effect (Waugh, Boyes, and Southon 1976), which modify the sensitivity of the energy barrier to the applied field. The impact of rolling-up is highlighted through a correlation of the C-map with zone line features in hitmaps. The region close to the (002) pole, where rolling-up is suspected to be significant, exhibits a different behavior in the variation of the energy barrier as a function of the field.

These observations are crucial for APT applications, as significant differences between crystallographic locations may alter local analysis conditions. Strong variations in the C-value could affect the curvature across the field of view and lead to spatial distortions in the reconstructed data. Additionally, it is important for controlling the field evaporation conditions in alloys, as a variation in the C-value could change the preferential field evaporation of one element over another from one crystallographic location to another, leading to compositional inaccuracies(Hatzoglou et al. 2020). It is worth noting that in the general case, the non-linearity of the energy barrier with the applied field should be considered. The C-value, which also measures the field sensitivity of the energy barrier, changes slightly with the base temperature.

Finally, FEELS method provides new insights into the physical process of field evaporation that are spatially resolved. However, it remains qualitative for the moment. Extracting local chemical information would require a systematic comparison of the measured C-value with density functional theory (DFT) field evaporation models, which are still in their infancy.

**Acknowledgements**

F.V. acknowledges the support of the French Agence Nationale de la Recherche (ANR), under grant ANR-21-CE42-0024 (project HiKEAP). Crystallographic dependence of Field Evaporation Energy Barrier in metals using Field Evaporation Energy Loss Spectroscopy mapping © 2024 by François Vurpillot is licensed under CC BY 4.0